\documentstyle[aps,prb,twocolumn,epsf]{revtex}

\begin{document}

\twocolumn[\hsize\textwidth\columnwidth\hsize\csname @twocolumnfalse\endcsname

\draft

\title{Order parameter fluctuation effects in 
 $d$-wave BCS superconductors}

\author{Hyok-Jon Kwon\cite{emailjon}}
\address{Department of Physics, University of Florida, Gainesville,
FL 32611-8440}

\date{\today}

\maketitle

\begin{abstract} 
We study  order parameter fluctuation effects in the superconducting
state as a possible precursor to the pseudogap phenomena.
 Using a low-energy effective theory 
in the $d$-wave BCS model, we self-consistently calculate the single-particle
properties. We find that the fluctuations reduce the spectral gap, and
the shape 
of the gap is deformed, showing reduced slope of the gap at the nodes.
We also show the angle-dependence of the quasiparticle lifetime
due to the fluctuations.
\end{abstract}
\vskip2pc ]
{\it Introduction}.---Despite much experimental evidence of 
pseudogap phenomena
in the underdoped cuprates, their microscopic mechanism is not 
understood.\cite{randeria97}
However, a pairing precursor as the origin of the pseudogap is one
prominent possibility.\cite{exp} 
An active endeavor has been  to incorporate 
strong pairing fluctuations to account for the pseudogap phenomena,
especially above the superconducting critical temperature
$T_c$.\cite{maly,varlamov}
In this report, we take a different tack 
and study the order parameter fluctuation effects in the superconducting 
state.
From a phenomenological standpoint, the pseudogap state can be considered
as a superconductor whose phase coherence is destroyed by strong phase
fluctuations whereas the gap is robust.\cite{Emery} 
Therefore, it will bare similarity 
to a superconducting state with strong order parameter phase and
amplitude fluctuations. 
Here we do not attempt to reproduce pseudogap phenomena since we
 study fluctuations in the weak-coupling BCS theory below $T_c$ and do not 
consider vortex 
pair unbinding transition, but
much of the qualitative trend is expected to pertain to the pseudogap state.
Using  the effective low-energy theory approach, 
we may describe the
problem with  relatively few physical parameters and  separate
the effect of order parameter phase and amplitude fluctuation 
effects.\cite{ours,xo}

We address the following two issues concerning the fluctuation effects.
First, we examine the effect of the order parameter fluctuations on the 
size of the  spectral gap below $T_c$. 
It has been shown that the fluctuations reduce the magnitude of the
order parameter and the critical temperature.\cite{varlamov,varlamov1,smith} 
In this paper we also show the reduction in the
spectral gap in a wide temperature range. For simplicity we do not the 
include Coulomb interaction
or disorder although they  would alter the form of the order parameter
fluctuations and further modify our result if properly 
included.\cite{varlamov,varlamov1}
 Secondly, we examine the angular variation of the fluctuation effect.
Angle-resolved photoemission spectroscopy (ARPES) data on underdoped cuprates
show that above 
or below $T_c$ the shape of the gap near the node significantly deviates 
from the simple $d$-wave shape.\cite{norman,mesot}
We show that this
may be due to the fluctuation of the phase rather than the amplitude
of the order parameter, and moreover that the amplitude fluctuation effect is
the strongest near the antinode. Also we discuss the angular variation
of the quasiparticle lifetime.

{\it Low-energy effective theory}.---We consider the weak-coupling mean-field 
BCS theory
in which the pairing potential gives
a $d$-wave order parameter which is effective only within the
momentum thickness $2\Lambda \ll k_F$ around the Fermi surface.
Then the effective fermion Hilbert space is a thin momentum shell of a
characteristic thickness $2\Lambda $ around the Fermi surface where
$\Lambda \ll k_F$.
For convenience, we coarse-grain the momentum shell into small boxes 
 and label them with an angular variable $\phi$.
The  effective action of the fermions and the order
parameter is
\begin{eqnarray}
S_{\rm eff} &=&\int d^2x \int_0^\beta d\tau \left[\sum_{\phi,\sigma}
c^{\dag}_\sigma(\phi ;{\bf x},\tau) 
\left(\partial_\tau +{\nabla^2 \over 2m} -\mu\right)
c_\sigma(\phi ;{\bf x},\tau)\right.  \nonumber \\
& &  + \sum_{\phi} \Psi^*({\bf x},\tau)
  w({\phi})
c_\downarrow (\phi ;{\bf x},\tau)
c_\uparrow (\phi+\pi ;{\bf x},\tau) +{\rm h.c.} \nonumber \\
& & \left. -{1\over g}\Psi^*({\bf x},\tau)\Psi({\bf x},\tau)
\right]~, 
\end{eqnarray}
where $\Psi$ is the superconducting order parameter introduced via
Hubbard-Stratonovich transformation to decouple the pairing interaction.
In the above we assume a  pairing potential of the form 
$V(\phi ,\phi^\prime )=g~w(\phi)w(\phi^\prime)$
where $w(\phi) = \cos 2\phi $ and $g<0$ which produces a $d$-wave
order parameter. It is understood that in writing 
$c_\sigma(\phi ;{\bf k},\tau)$, the momentum $\bf k$ lives only inside 
the small box labeled by the angular variable $\phi $ near the Fermi
surface.
In order to explicitly separate the order parameter phase and amplitude
degrees of freedom, we re-express $\Psi({\bf x},\tau)=\Delta({\bf x},\tau)
e^{i\theta({\bf x},\tau)}$ where $\Delta({\bf x},\tau)$ takes a real value.
In the mean-field approximation, we replace $\Delta({\bf x},\tau)$ 
with $\Delta_0$ and obtain a d-wave gap $\Delta(\phi)=\Delta_0
\cos 2\phi $ using the following self-consistent gap equation:
\begin{equation}
{1\over |g|} = T\sum_{\omega}\sum_{\phi ,{\bf k}}
{w^2(\phi) \over \omega ^2 +\xi_{\bf k}^2 +\Delta^2(\phi)}~,
\label{gapeq}
\end{equation}
where $\xi_{\bf k}=k^2/2m -\mu $. The momentum summation above is
constrained by the condition $|\xi_{\bf k}| < v_F \Lambda $.
 Here we consider the fluctuation around the mean-field
value and  re-express $\Delta({\bf x},\tau) = \Delta_0+d({\bf x},\tau)$.
Then we perform a gauge transformation 
$ \psi_\sigma({\bf x},\tau) = c_\sigma({\bf x},\tau)  
e^{-i\theta ({\bf x},\tau)/2}$, to couple the phase fields to the
fermions explicitly.
 
The resulting effective action is expressed in terms of the Nambu 
spinor notation,
$\hat{\psi} = (\psi_{\uparrow}, \psi^{\dag}_{\downarrow}) $, as
$S_{\rm eff}= S_0 + S_I$,
with 
\begin{eqnarray}
S_0&=& T\sum_{\omega}
\sum_{\phi ,{\bf k}} \hat{\psi}^{\dag} \hat{G}_0^{-1} \hat{\psi} 
     \\ \nonumber
 && +\int_0^\beta d\tau 
\int d^2x \big{\{} \, {n_f \over 8m} \,[\nabla \theta ({\bf x},\tau )]^2
+{1\over g} [d({\bf x},\tau)]^2 \big{\}}
\label{Act0}
\end{eqnarray}
and
\begin{eqnarray}
S_I &=& T\sum_\omega T\sum_\nu \sum_{\phi, \bf k,q} 
\hat{\psi}^{\dag}(\phi ; {\bf k},\omega)\Big{\{}
{1\over 2}[-\nu +i{\bf v}_F(\phi)\cdot {\bf q}
]\theta({\bf q},\nu) \nonumber \\
&& +w(\phi)
\hat{d}({\bf q},\nu)\Big{\}}\hat{\psi}(\phi ; {\bf k-q},\omega -\nu)
~.
\label{ActI}
\end{eqnarray}
Here
\begin{equation}
 \hat{G}_0^{-1} = \left( 
\begin{array}{cc} 
i\omega - \xi_{\bf k} & \Delta_0 w(\phi) \\
\Delta_0 w(\phi) & i\omega + \xi_{\bf k} \\
\end{array}
 \right),
\end{equation}
and $ \hat{d}(\nu , {\bf q}) =\hat{\sigma}_x~{d}(\nu , {\bf q})$,
with $\hat{G}_0$ the bare Green's function for the neutral fermions.
In the above, we approximate $\xi_{\bf k}=k^2/2m -\mu $ as
$\xi_{\bf k} \approx v_F( |{\bf k}| -k_F)$ and ${\bf v}_F(\phi)$
is the Fermi velocity in the $\phi$ direction.

In building this effective theory 
we have not considered  vortex pair unbinding which 
leads to the Kosterlitz-Thouless transition.
This is justified well outside the fluctuation regime. We also assume that
we are in the temperature range where the BCS mean-field theory is justified, 
namely, that $\delta T/T_c \gg \Delta_0/E_F $
in a two dimensional clean superconductor.\cite{AL}
One thing we observe from the form of the effective theory is that
the strength of the coupling between fermions and the amplitude 
fluctuations has an
angle-dependence; the amplitude fluctuation effect 
is suppressed near
the gap nodes, as is evident in Eq. (\ref{selfE0}) where the
self-energy correction is multiplied by a factor of $w^2(\phi)$. 
The strength of the coupling to the phase fluctuations is not suppressed 
at the node, however.

In studying the finite temperature superconducting to normal state
transition, it should be sufficient to consider only the static
fluctuations of the phase, and we may suppress the time-dependence
in $\theta $ and $d$ and retain only the spatial fluctuations.
Eq. (\ref{ActI}) is then modified as
\begin{eqnarray}
S_I &=& T\sum_\omega \sum_{\phi, \bf k,q} 
\hat{\psi}^{\dag}(\phi ;{\bf k},\omega)\Big{\{}
{1\over 2}i{\bf v}_F(\phi)\cdot {\bf q}
\theta({\bf q}) \nonumber \\
&& +w(\phi)
\hat{d}({\bf q})\Big{\}}\hat{\psi}(\phi ; {\bf k-q},\omega)
~.
\label{ActIm}
\end{eqnarray}
The simplified form above, however, does not produce reliable results
near zero temperature.

{\it Fermion single-particle properties}.---In evaluating the quasiparticle 
self-energy by perturbative expansions,
we take advantage
of the fact that the effective theory resides in the thin shell around
the large Fermi surface and select the diagrams which are of leading
order in 
$\Lambda /k_F$,
which amounts to summing over the ring diagrams
in calculating the order parameter correlation function.
The fermion self-energy can be obtained self-consistently from the
Dyson equation.

We first evaluate the correlation functions of the amplitude
fluctuations: 
\begin{eqnarray}
\langle d({\bf q})~d({-\bf q})\rangle _{\rm ring} &=&
{g~T\over 1 + {g}\Pi_{dd}({\bf q},0)} ~,
\label{ddcor}
\end{eqnarray}
where
\begin{eqnarray}
\Pi_{dd}({\bf q},\nu) &=& 
{1\over 2}T\sum_{\omega }\sum_{\phi , \bf k}
w^2(\phi) {\rm Tr} \left[ \hat{G}_0(\phi ;{\bf k},\omega)
\hat{\sigma}_x \right. \nonumber \\
&& \left. \times \hat{G}_0(\phi ;{\bf k+q},\omega+\nu)\hat{\sigma}_x
\right] ~.
\label{Pdd}
\end{eqnarray}
From Eq. (\ref{ddcor}) we obtain
$\langle d({\bf q})~d({-\bf q})\rangle _{\rm ring}={T/( a+b~q^2)}$
where
the temperature-dependent coefficients $a$ and $b$ can be evaluated
by carefully expanding $\Pi_{dd}$ in $\bf q$ from Eq. (\ref{Pdd}).
If we only consider the spatial fluctuations in the order parameter,
the $\langle d~\theta\rangle $ terms are zero
in the Gaussian approximation.
Also the phase fluctuation has the following well-known correlation
function\cite{ours}
\begin{equation}
\langle \theta({\bf q})\theta({\bf -q})\rangle _{\rm ring}=
{4mT \over n_s(T)~q^2}~,
\end{equation}
where $n_s(T)$ is the superfluid density at temperature $T$.

Now we can determine the quasiparticle self-energy correction using
the self-consistent Dyson equation, neglecting the vertex corrections:
\begin{eqnarray}
\hat{\Sigma}(\phi ;{\bf k}, \omega) &\approx &
\sum_{\bf q} \left\{ {1\over 4} \left[ {\bf v}_F(\phi)\cdot {\bf q} \right] ^2
\langle \theta({\bf q}) \theta({\bf - q})\rangle _{\rm ring}
\right. \\ \nonumber 
&&  + w^2(\phi) \langle d({\bf q}) d({\bf - q}) \rangle _{\rm ring}
\bigg{\}} \hat{G}(\phi ;{\bf k-q}, \omega -\nu)~,
\label{selfE0}
\end{eqnarray}
where $\hat{G}$ is the full fermion Green's function, given
self-consistently by $\hat{G}^{-1} = \hat{G}^{-1}_0 - \hat{\Sigma}$.
In general the self-energy has both a momentum and frequency dependence,
but we focus on the behavior of the self-energy near the Fermi
surface, assuming that it varies smoothly near the Fermi surface. Therefore 
we neglect the $\xi_{\bf k}$-dependence so that the only 
momentum dependence is  through the angle $\phi$ on the
Fermi surface.
Then we can approximately obtain the self-energy:
\begin{eqnarray}
\hat{\Sigma}(\phi ,\omega) &\approx & \left\{
{4mT\over n_s(T)}\,{1\over 16\pi} \ln \left[
{\Lambda ^2 + \tilde{\Delta}^2(\phi)+\tilde{\omega} ^2 \over 
\tilde{\Delta}^2(\phi)+\tilde{\omega} ^2} \right] \right. \nonumber \\ 
&& +\left. \sum_{\bf q} {T~w^2(\phi)\over a+b~q^2}~{1\over
\tilde{\omega} ^2 +({\bf v}_F(\phi)\cdot {\bf q})^2 +\tilde{\Delta}^2(\phi)}
\right\} \nonumber \\ 
&& \times \left( 
\begin{array}{cc} 
-i\tilde{\omega} & \tilde{\Delta}(\phi) \\
\tilde{\Delta}(\phi) & -i\tilde{\omega} \\
\end{array} \right),
\label{selfE}
\end{eqnarray}
where $\tilde{\omega}$ and $\tilde{\Delta}$ can be calculated 
self-consistently by
$\hat{G}^{-1}(\phi ;{\bf k},\omega) = \hat{G}_0^{-1}(\phi ;{\bf k},\omega)
-\hat{\Sigma}(\phi, \omega)$.

From the self-energy obtained in Eq. (\ref{selfE}), by analytically
continuing 
the frequency $i\omega \rightarrow
\omega +i\eta $,
we can calculate various single-particle
properties such as density of states (DOS), spectral functions, 
and single-particle
scattering rates. In this paper we focus on DOS since
it is gauge-invariant and measurable via the tunneling spectroscopy\cite{sts}
 or  the momentum-integrated ARPES data\cite{photo}.
We are especially interested in the
angle-resolved DOS:
\begin{equation}
N(\phi ,\omega) = -{1\over \pi} {\rm Im}\int d\xi _{\bf k}~ {\rm Tr}~
\hat{G}(\phi ;{\bf k},\omega)~,
\end{equation}
as it gives information about the angular variation of the fluctuation
effect.

Throughout this paper, we set the relevant energy scales 
$\Lambda \approx 10\Delta _0(T=0)/v_F $ and $E_F \approx 5 \Lambda v_F$ so that
we are well in the BCS weak-coupling regime; these relative energy scales 
give much
stronger pairing strength than ordinary superconductors but significantly
weaker than the cuprates.
With the energy scales so chosen, we may estimate the regime of the
validity of the mean-field theory. If we apply the Ginzburg criterion,
 namely, $|\Delta_0(T)|^2 \gg \langle d({\bf x})~d({\bf x})\rangle $,
which may be estimated from Eq. (\ref{ddcor}),
the mean-field theory breaks down only near $T/T_c \sim 0.98 $.
Therefore, the BCS framework is reliable in most of the temperature
range that we consider. 

\begin{figure}
\epsfxsize=6.5cm \epsfbox{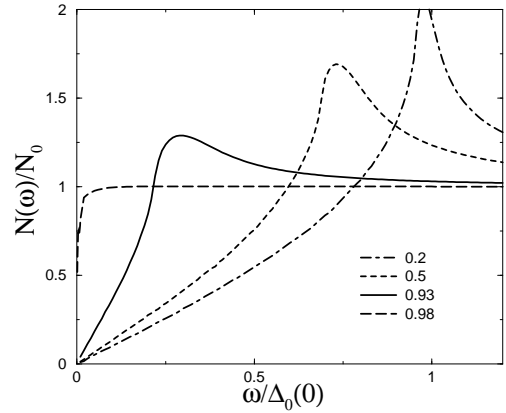}
\bigskip
\caption{Total density of states ($N(\omega)/N_0$) as a function of
energy 
at four temperatures ($T/T_c = 0.2,~0.5,~0.93,$ and 0.98).
 }
\label{TDOS}
\end{figure}

\begin{figure}
\epsfxsize=6.5cm \epsfbox{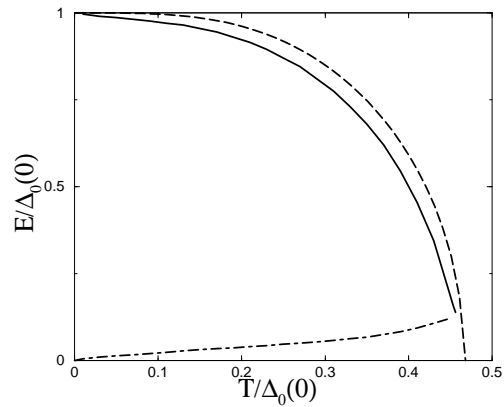}
\bigskip
\caption{The comparison of the density of states peak to the mean-field
gap magnitude. The solid line is the total density of states peak position
and the long dashed line is the mean-field gap. The dot-dashed line
is the difference between the two quantities.
}
\label{peak}
\end{figure}

In Fig. \ref{TDOS} we show the total DOS. As the temperature
increases, the DOS peak is widely smeared. Very close to 
$T_c$, the DOS peak has almost disappeared and the spectral gap is only 
manifested by the depletion in the DOS around $\omega =0$ as compared to
the normal state DOS.
Figure \ref{peak} shows the DOS peak position as a function of temperature;
in most of the temperature range, we can interpret the peak position roughly 
as the spectral gap.
This figure shows that
the spectral gap is reduced in a wide temperature range due to the order 
parameter fluctuations.
Near  $T/T_c\approx 0.98$, the DOS peak structure has almost disappeared, 
and the DOS maxima do not have a meaning as the spectral gap. Therefore,
we need to
estimate the size of the spectral gap from the width of the DOS depletion
in this case.
It is difficult to study the evolution of the spectral gap  
through $T_c$ in this framework because the mean-field theory breaks down 
sufficiently close to $T_c$ as discussed above. 
Also the result near zero temperature is not reliable
due to the negligence of the time-dependence of the fluctuations.

\begin{figure}
\epsfxsize=6.5cm \epsfbox{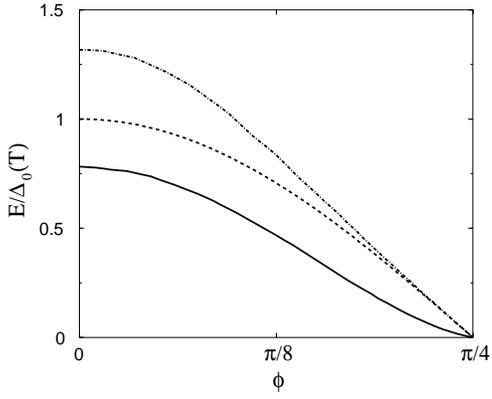}
\bigskip
\caption{The peak points of the angle-resolved density of states 
(solid line) vs the mean-field gap (dashed line) at
 $T/T_c = 0.93$. The dot-dashed line is the  DOS peak curve evaluated
with only the order parameter amplitude fluctuations.
}
\label{dp45}
\end{figure}

The angle-dependent DOS peak near $T_c$ is shown in Fig. 
\ref{dp45}. 
At $T\ll T_c$, the DOS peak contour follows the $d$-wave shape.
As the temperature approaches $T_c$, we observe that the DOS
is widely smeared to  low-energy states especially near the antinode
($\phi=0$).
Figure \ref{dp45} shows that 
the shape of the angle-resolved gap (DOS peak curve) is deformed
from the original $d$-wave shape near the node. We argue that the 
downward bending of the DOS peak curve near the node ($\phi = \pi/4$) 
is due to the phase fluctuations since the amplitude fluctuations alone
do not cause the downward bending as illustrated in the same figure.
We find that the angle-dependence of the spectral gap near the node is
\begin{eqnarray}
|\tilde{\Delta}(\phi)| &\approx & \Delta_0(T)  |\cos(2\phi)| \\
&&~\times \left\{ 1-{mT\over2\pi n_s(T)} 
\ln \left[{\Lambda\over \Delta_0(T)|\cos(2\phi)|}
\right] \right\}~, \nonumber
\end{eqnarray}
and the slope of the gap near the node is reduced.
It has similarity to the shape of the gap obtained by ARPES  on 
underdoped $\rm Bi_2Sr_2CaCu_2O_{8+\delta}$ (Bi2212) in the superconducting 
state,\cite{mesot} although its microscopic origin is not well understood.

\begin{figure}
\epsfxsize=6.5cm \epsfbox{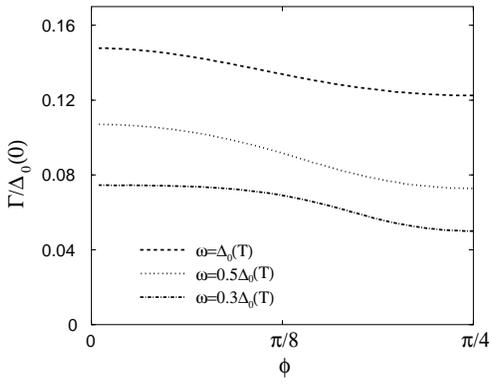}
\bigskip
\caption{
The angular variation of the
quasiparticle scattering rate ($\Gamma (\omega) / \Delta_0(0) $ )
evaluated at $\omega/\Delta_0(T) =1, ~0.5,$ and 0.3 
at $T/T_c = 0.93$. }
\label{rate}
\end{figure}

Figure \ref{rate} shows angular variation of the scattering rate due to the 
order parameter fluctuation.
We observe that the maximum scattering rate occurs near the antinode
and also that the rate decreases as one approaches the node.
This variation is due to the angular dependence of the order parameter
magnitude fluctuations.
This feature may contribute to the anisotropy of the quasiparticle
scattering rate in cuprate  superconductors.\cite{photo} 


{\it Discussions and Conclusions}.---Here we discuss the qualitative
effects of amplitude and phase fluctuations.
In Fig. 3, we find that if we omit the phase fluctuation effect,
the apparent gap is enhanced. This is because the amplitude
fluctuations tend to increase the gap magnitude. This can be understood
from the Fermi liquid reference frame as following:
The fermion self-energy correction due to the amplitude fluctuations
can be roughly estimated as
\begin{eqnarray}
\Sigma ({\bf p}, \omega) & \approx & -\int d \nu d^2 q
G({\bf p -q},\omega -\nu )~\langle \Delta ({\bf q},\nu)\Delta ({-\bf q},-\nu)
\rangle \nonumber \\
&\approx & {1\over i\omega + \xi_p} \langle \Delta (x) \Delta (x)\rangle ~,
\end{eqnarray}
and therefore the effective spectral gap is enhanced by the fluctuations as
$|\Delta_{\rm eff}|^2 = \langle \Delta (x) \Delta (x)\rangle
=|\Delta_0|^2+\langle \delta\Delta (x) \delta\Delta (x)\rangle$.

The effect of phase fluctuation can be considered as a Doppler shift in
the fermionic spectrum by ${\bf k}_F\cdot {\bf v}_s$ where ${\bf v}_s \sim
\nabla \theta /m$. Due to thermally fluctuating superfluid velocity, 
the DOS near the gap
is now shifted
since the energy levels at the gap nodes are enhanced.
As a result, more states would be occupied near the nodes, and hence
decrease in the slope of the gap nodes as shown in Fig. 3.
On including the vortex pair unbinding transition of the BKT type, 
which gives stronger phase fluctuation effects, we can obtain a
Fermi arc-like phenomenon.\cite{ours,franz98} 

In order to obtain the correct effect on the size of the spectral gap,
both the phase and amplitude fluctuations have to be self-consistently
taken into account. The total effect is reduction in the spectral gap as 
shown in Fig. 2 and 3. 
However, the spectral gap
may not be equal to the order parameter magnitude especially if the
fluctuation is strong,\cite{kosztin} and hence more careful
study is needed to separate these two quantities.

The results presented in this report are equally well applicable to any 
unconventional superconducting symmetry. In principle, any superconductor
would have a window of temperatures near $T_c$ where such fluctuations are
visible, depending on the pairing strength and the superfluid density.
In case of  underdoped cuprates,
however, it would be essential to include the effect of vortex pair 
unbinding in the pseudogap state, due to the small superfluid density. 
The result of this report may nevertheless pertain to its superconducting 
state. Indeed, a recent
observation of the deformed gap shape
in Bi2212,\cite{mesot} which is only observed in
underdoped regime,
may be related to the order parameter fluctuation effects,
considering that the phase fluctuations are more important in underdoped
cuprates due to reduced superfluid density. Since the 
microscopic origin of this deformation is not understood, further experimental
investigation on the temperature variation of the gap anisotrpy would
be desirable.

Some of the above features are shown to be shared by other 
theoretical results in the normal state counterpart.
For instance,
the form of the density of states obtained above is similar to that above $T_c$
when the Gaussian pairing fluctuations are incorporated.\cite{old}
Also a similar but much more pronounced deformation of the $d$-wave spectral 
gap was obtained above $T_c$ using a self-consistent conserving 
approximation.\cite{jan}

%
%
%
%

The author thanks Alan Dorsey and Rob Wickham for helpful discussions and
comments.
This work was supported  by the 
National High Magnetic Field Laboratory and by NSF grant DMR 96-28926.  

\end{document}